# Processing and study of the composite CdS/Bi-Pb-Sr-Ca-Cu-O

E. Díaz-Valdés[1], G. S. Contreras-Puente[1], N. Campos-Rivera[1],
C. Falcony-Guajardo[2], and R. Baquero[2].

[1]Departamento de Física, ESFM-IPN, México D.F., México
[2]Departamento de Física, CINVESTAV-IPN, México D. F., México
Corresponding author: e-mail: elviadv@esfm.ipn.mx

*Abstract* – **We have fabricated and characterized samples of the superconducting-semiconducting Bi-Pb-Sb-Sr-Ca-Cu-O/CdS composite. Nano-size particles of CdS were deposited and introduced into the porosities of the Bi-Pb-Sb-Sr-Ca-Cu-O material by the spray pyrolysis technique. The morphology and hollow size in the porous superconducting material as well as the grain size in CdS and the morphology of the surface of the composite were obtained by Scanning Electron Microscopy. We obtained the critical superconducting temperature of both the Bi-Pb-Sb-Sr-Ca-Cu-O and the Bi-Pb-Sb-Sr-Ca-Cu-O/CdS composite measuring the resistivity. Both show a metallic behaviour just above the superconducting transition. For the superconductor alone, resistivity starts falling at $T_{c,on\,sup}$ = 99,9 K and reaches zero at $T_{c,sup}$=76,3 K. The behaviour of the composite is different. The transition starts at $T_{c,on\,comp}$ = 65,3 K and reaches zero resistance at $T_{c,comp}$ = 56,5 K. This seems to indicate that the semiconductor penetrates the whole superconducting Bi-Pb-Sb-Sr-Ca-Cu-O sample so that there is no region of pure superconducting material left. Since the materials do not actually mix (see text) the behaviour might be attributable to the interface. Also the resistivity curves present a very interesting feature, i.e., below the temperature at which the composite attains zero-resistivity, a re-entrant behaviour manifests itself and a finite resistivity peak appears. It increases to a certain value to drop back to zero at some temperature below. We comment further on this feature in the text.**

## 1 Introduction

Since the discovery of high-$T_c$ superconductivity [Berdnoz and Muller, 1986; Fung and Kwok, 1996], a huge amount of new materials were fabricated. New ceramics and composites have been processed. Examples of these last are the hetero-structured, semiconducting and superconducting composites [Petrov et al., 2001; Tang et al., 2005]. The main objective of these composites is to take advantage of some of the properties of both materials. Even though, the mechanism that leads to the superconducting phase transition in the high-Tc compounds is not actually known, experimental results show that the essential physical manifestations are the same as in conventional superconductivity. In this work, we have processed a superconducting-semiconducting composite $Bi_{1.6}Pb_{0.4}Sb_{0.1}Sr_2Ca_2Cu_3O_\delta$/CdS material that shows a good incorporation of CdS into the

Bi-compound. The composite presents new and interesting effects. We present here its characterization as we have worked it out so far. The composite presents a re-entrant behavior in the resistivity as a function of temperature that might be used as a thermal switch. (see below)

## 2 Experimental

Superconducting pellets were prepared with $Bi_{1.6}Pb_{0.4}Sb_{0.1}Sr_2Ca_2Cu_3O_\delta$ nominal composition, by the solid state reaction method. Six thermal treatments between 400ºC and 860ºC and from 10 hours to 30 days were applied following the usual procedures in the literature [Perea, 2009]. CdS was prepared by the co-precipitation technique from cadmium nitrate and sodium sulphur. The morphology and the hollow size of superconducting porous material, as well as, the size of the grains in the semiconducting material were obtained by Scanning Electron Microscopy in the Sirion model FEI microscope. Grown phases in the semiconducting and superconducting material were identified by X-ray diffraction in the D8 Focus Bruker diffractometer. A CdS solution 1,43 x $10^{-4}$ kg/L was prepared by dissolving it in $H_2SO_4$ (5% V). This solution was deposited on the porous superconducting Bi-Pb-Sb-Sr-Ca-Cu-O material by spray pyrolysis . For deposition, the superconducting pellet was on a bed of liquid tin at 300ºC. The solution is atomized and the aerosol that is decomposed with temperature is guided to the pellet with an air flux. The deposition time was 30 min.

## 3 Results

### 3.1 X-ray diffraction

CdS usually grows in the wurzite crystal structure but it can also be grown in the cubic phase. We have grown cubic face-centred CdS with a lattice parameter $a = 5.83$ Å. Superconducting $Bi_{2-x}Pb_xSr_2Ca_2Cu_3O_\delta$ was grown in the orthorhombic crystal structure with lattice parameters $a = 5.407$ Å, $b = 5.416$ Å and $c = 37.118$ Å. The X-ray diffraction pattern of the composite material shows a mixture of $Bi_{2-x}Pb_xSr_2Ca_2Cu_3O_\delta$, $Bi_2(Sr_{1.8}Bi_{0.2})CuO_{6,16}$ and CdS phases. Figure 1 shows our X-ray spectra.

### 3.2 Electrical Measurements

Figure 2 shows our measured resistivity as a function of temperature for the superconductor alone and for the composite material. The superconducting material alone presents metallic behaviour at temperatures just above the transition. The resistivity starts falling at $T_{c,on\ sup} = 99.93$ K and achieves zero resistance at $T_{c,0\ sup} = 76.29$ K. So actually, the width of the superconducting transition, $\Delta T = 23.64$ K, amounts to roughly a 30% of the critical temperature of the material which might point to a crystal with some imperfections in the structure. It is therefore interesting that the composite material presents a much narrower transition of about 9 K only. This might mean that the transition takes place at a quite clean interface between the semiconductor and the superconductor. No traces of the host material were found since the transition at a higher Tc of the clean superconductor did not show in the resistivity measurements in this case. The composite showed a metallic behaviour just above the transition to the superconducting state. At $T_{c,on\ comp} = 65.3$ K, the resistivity began to fall and zero resistance was attained at $T_{c,0\,comp}$ = 56.51 K, therefore showing a superconducting transition width of $\Delta T$=8.79 K. The width in this case is quite smaller that the one observed for the superconductor alone, as we stated above.

A very peculiar behaviour was observed for the composite. It appears as an intrinsic property of the material as we have checked it so far. It turns out that a reproducible re-entrant behaviour takes place below the critical temperature. A turn up of the resistivity curve to non-zero values was measured which means that a transition back to the normal state took place at that point, leaving therefore the superconducting state. At a lower temperature, resistivity falls continuously to zero and the superconducting state is re-establish. Below the critical temperature, as the re-entrant behaviour appears, the resistivity peak grows to a maximum at T= 47.27 K and falls down to zero at 34.72 K so that a new transition back to the superconducting state takes place. We offer a speculation on the origin of the behavior of this peak in the text below.

*3.3 Scanning Electron Microscopy*

Figure 3 shows the CdS micrograph obtained with secondary electrons and magnified 25000X. CdS particles smaller than 0.2 $\square$m can be observed.

Figure 4 shows the superconducting surface micrograph obtained with secondary electrons and magnified 500X. It can be observed a hollow size around 25 $\square$ m, indicating that CdS particles can be embedded into the hollows of the superconductor.

Figures 5 and 6 show composite micrographs obtained with secondary electrons and magnified 500X and 4000X respectively. In these figures it can be appreciated that the superconducting surface was covered by the CdS particles showing its introduction into the hollows in the superconducting Bi-compound.

**4 Discussion**

X-ray diffraction profiles obtained from superconducting, semiconducting and composite material indicate that the last one is immersed into the superconducting material without changing its structure.

The micrographs show that the CdS particles are into hollows and styked on the walls of the superconducting material. As a result we can say that semiconductor material is not only the on surface but also inside the superconductor forming therefore a true composite material.

Electrical measurements show important changes in the superconducting behaviour before and after deposition. A feature to be noticed concerns the shape of the transition of the composite to the superconducting phase. Two points are interesting:

$$T_{c,ons,super} > T_{c,on,comp} \tag{1}$$

$$\Delta T_{super} > \Delta T_{comp} \tag{2}$$

In the composite, clean interfaces between the superconductor and the semiconductor form, as it can be judged from the fact that the transition to the superconducting state is relatively sharp, sharper than in the clean superconducting sample. The interfaces might have a metallic character even on the semiconducting side of the interface on a few atomic layers. This proximity effect was found before at the interface YBCO7/GaAs [García et al., 2008], for example. In this interface the two atomic layers that constitute the interface on the GaAs side, were found to be metallic. Therefore, the electrons that constitute the Cooper pairs as well as the intermediate boson (whatever it is) change their dispersion relations which, in general, might not be the same as the ones in the bulk materials. Also the specific states that might build up inside the CdS nano-structures can contribute to specific characteristics of the superconducting state. These details are under study.

**On the re-entrant behavior.** Another remarkable result in the composite curve resistivity vs. temperature is its re-entrant behavior illustrated in Fig. 2, and mentioned above in the text. Below the critical temperature, at T= 47.27K, the resistivity goes back to non-zero values, cracking the superconductive phase. Then, on decreasing temperature, a maximum at 42.36K is formed and the resistivity decreases back to zero at T=34.72K. The fact that the resistivity rises as the temperature falls can be associated to a known behavior of certain superconductors that are near a dielectric or a semiconducting transition. First, on lowering the temperature, the system behaves as a metal and the resistivity decreases steadily. Then, from a certain temperature down, as the conduction band empties, the resistivity increases up to a certain value and then, at a certain point, the phase transition occurs because the pairing energy accumulated in the Cooper pairs, makes superconductivity the ground state. It is therefore a question of energy balance between the energy lost by occupying the lowest available electronic states by empting the conduction band, on one side, and the lost of energy that the building of the Cooper pairs represents, on the other. This behavior is known but the new feature here is that the resistivity does go to zero for a certain range of temperature before presenting the re-entrant behavior. The usual case is that resistivity does present a peak of the same kind but in a region previous to the critical temperature and without going to zero. We think that the same kind of energy balance takes place in our case. The question is what kind of changes in the sample brings back this kind of semiconducting behavior? We try to answer this question in what follows

As we show in Fig. 6, the composite superficial layer is a set of superconducting and semiconducting particles embedded on each other. If the interfaces are the ones mostly responsible for superconductivity, then they built a zero-resistivity path by successive Josephson tunneling from one to the next interface. This delicately build zero-resistivity path can change as a consequence of different factors. One possibility is that when the current flows into the composite material, a percolation phenomenon, or electrical transport by percolation, could be produced [Grassberger, 1991; Rivera, 2003]. We are investigating further the details of this feature,

## 5 Conclusions

We have, in this work, fabricated and characterized the semiconductor/superconductor composite material **CdS/Bi-Pb-Sr-Ca-Cu-O**. The composite presents regions of the superconductor and the semiconductor material embedded on each other. The composite as a whole, behaves in the normal state as a semiconductor. This conclusion was reached from a simple model that assumes that the sample can be modeled approximately by a parallel circuit.   A very interesting result is that the resistivity curve vs temperature presents a re-entrant behavior. At T= 47.27K, the resistivity goes back to non-zero values, cracking the superconductive phase. Then, on decreasing the temperature, a maximum at 42.36K is formed and the resistivity decreases back to zero at T=34.72K. We speculate (see the text for more details) that the zero-resistivity path breaks as the consequence of a delicate balance between the condensation energy of the Cooper pairs and the lost in energy of the system by occuoying the lowest available electronic states (the semiconducting behavior) [Baquero, 2006]. The peak could be used as a thermal switch. We are at the present studying further the characteristics of this peak, in particular, its behavior under a magnetic field. We will present these results in a following publication.

**Figure 1** The X-ray diffraction pattern of the composite material shows a mixture of $Bi_{2-x}Pb_xSr_2Ca_2Cu_3O_\delta$, $Bi_2(Sr_{1.8}Bi_{0.2})CuO_{6.16}$ and CdS phases. Figure 1 shows our X-ray spectra

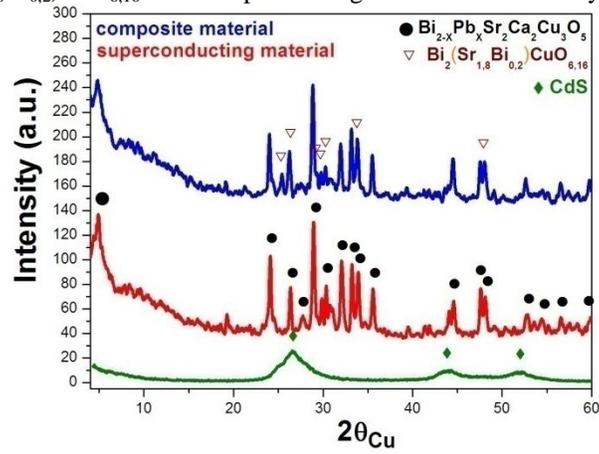

**Figure 2** Bi-Pb-Sr-Ca-Cu-O superconductor and CdS/Bi-Pb-Sr-Ca-Cu-O composite electrical behaviour.

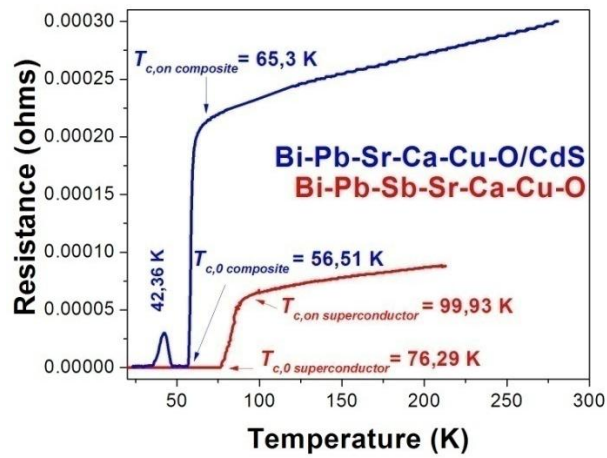

**Figure 3** CdS micrograph.

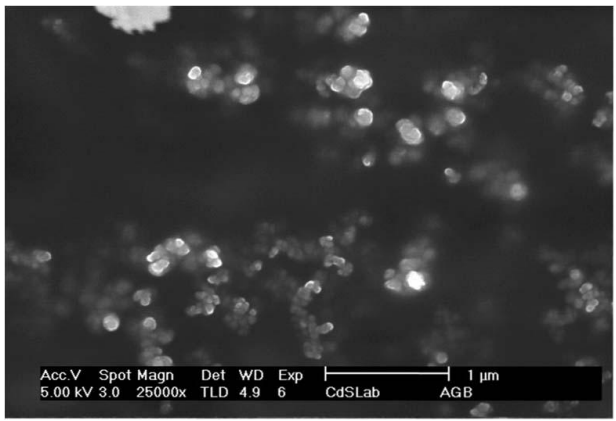

**Figure 4** Bi-Pb-Sr-Ca-Cu-O superconductor micrograph obtained with secondary electrons.

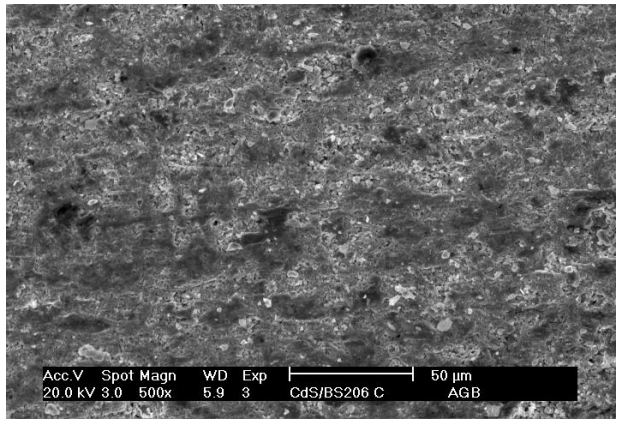

**Figure 5** Bi-Pb-Sr-Ca-Cu-O/CdS composite micrograph obtained with secondary electrons and magnified 500X.

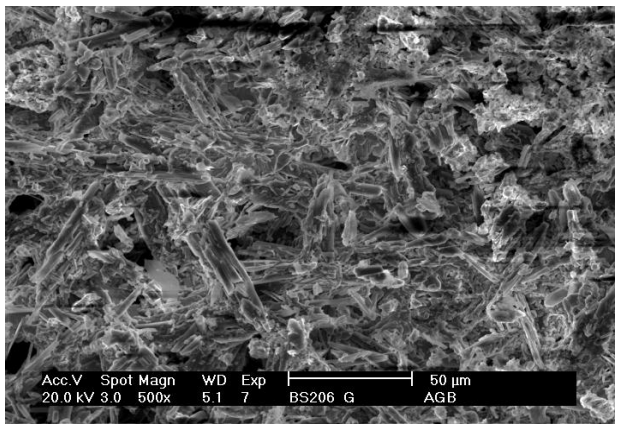

**Figure 6** Bi-Pb-Sr-Ca-Cu-O/CdS composite micrograph obtained with secondary electrons and magnified 4000X.

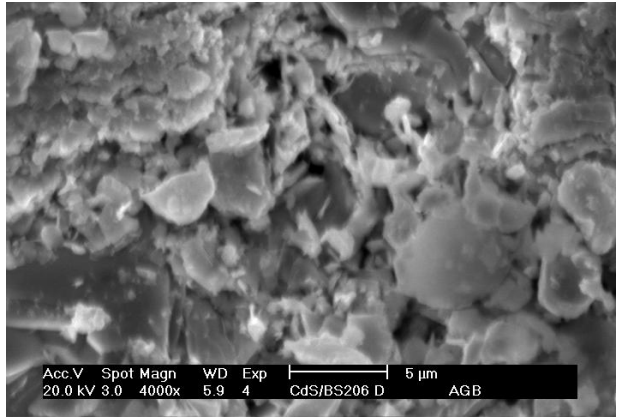